\documentstyle[prd,aps,floats,epsf,color]{revtex}
\begin{document}
\baselineskip=12pt
\twocolumn[\hsize\textwidth\columnwidth\hsize\csname@twocolumnfalse\endcsname

\title{$f(R)$ Gravity: From the Pioneer Anomaly to the Cosmic Acceleration}

\author{Reza Saffari$^1$, Sohrab Rahvar$^{2}$}
\address{$^1$ Institute for Advanced Studies in Basic Sciences, P.O.
Box 45195--1159, Zanjan, Iran}
\address{$^2$ Department of Physics, Sharif University of Technology,
P.O.Box 11365-–9161, Tehran, Iran}

\vskip 1cm
\maketitle

\begin{abstract}
We use metric formalism in $f(R)$ modified gravity to study the
dynamics of various systems from the solar system to the
cosmological scale. we assume an ansatz for the derivative of action
as a function of distance and describe the Pioneer anomaly and the
flat rotation curve of the spiral galaxies. Having the asymptotic
behavior of action, we propose the action of $f(R) = (R + \Lambda)(1
+ \ln(R/R_c)/(R/R_0 + 2/\alpha))$ where in galactic and solar system
scales it can recover our desired form. The vacuum solution of this
action also results in a positive late time acceleration for the
universe. We fix the parameters of this model, comparing with the
Pioneer anomaly, rotation curve of spiral galaxies and Supernova
Type Ia gold sample data.
\end{abstract}
\newpage
\vspace{1cm}

]

\section{\label{sec:level1}Introduction}
Recent observations of the Supernova Type Ia (SNIa) and Cosmic
Microwave Background (CMB) radiation indicate that universe is under
positive accelerating expansion ~\cite{wmap3a}. This accelerating
expansion is one of the important puzzles of the contemporary
physics. A nonzero vacuum energy can derive universe to accelerate
however one can ask why it is non-zero and why it is so
small\cite{sah99}. Adding a simple cosmological constant term to the
Einstein equations can also accelerate universe. However, the
problem with the cosmological constant is that why the energy
density of matter and cosmological constant are in the same order at
the present time?

A varying dark energy model can partially solve this problem in
which the density of dark energy traces the density of matter from
the early universe to the present time. Modified gravity can also
provide an effective time varying equation of state. In these models
the Einstein-Hilbert action is replaced with a generic form of
$f(R)$ gravity \cite{carr04}. In addition to the late time cosmic
expansion, early inflationary era also can be achieved by an extra
term to the action, as adding a cubic term to $1/R$ gravity model
\cite{sot06}. Modifying action not only affects the dynamics of
universe, it can also alter the dynamics at the galactic or solar
system scales.

There are two main approaches of metric and Palatini formalism to
extract the field equations from the action. Considering the non
Levi-Civita connection associating to the manifold, we can take the
connection and the metric as the independence geometrical
quantities. Varying the action with respect to these two parameters
(so-called Patatini formalism) results in the field equations
\cite{VolPhys,OlmoPhys2}. On the other hand in the metric formalism
the connection is the Levi-Civita connection and we do variation of
action with respect to the metric to derive modified gravity field
equations. The advantage of the Palatini formalism is that the field
equations are second order differential equations similar to the
other parts of the physics.

One of the interesting issues in $f(R)$ gravity is studying the
spherically symmetry solutions. In the case of Palatini formalism
the solution is Schwarzschild--de'sitter metric with an effective
cosmological constant.
However in the metric formalism the solution of non
Einstein-Hilbert action suffers from a low-mass equivalent scalar
field that is incompatible with Solar System tests of general
relativity, as long as the scalar field propagates over Solar System
scales \cite{kam06,chi07}.
One of the solutions to evade the solar system tests is using action
in such a way that reduces to the Einstein-Hilbert action in the low
curvature regime at the solar system and at the cosmological regime
acts as an effective cosmological constant \cite{star07,star07-2}.
The other problem in this issue is the consistency of the
spherically symmetric solutions in $f(R)$ gravity \cite{saf07} that
will be addressed in this paper.

In this work, we try to extract an appropriate action for the
modified gravity through the inverse solution. This method has been
applied in the previous works both in the galactic \cite{sob07} and
cosmological scales \cite{rah_sob}. Here we extend the previous
works to the solar system scale, studying anomalous in the Pioneer
acceleration and obtain the appropriate action in the solar system
scale. On the other hand following the method proposed by
Capozziello et al. \cite{capo07} we extract an appropriate action to
provide a flat rotation curve in the spiral galaxies. Both the solar
system and the galactic scale solutions are consistent with the
modified gravity field equations in the first order of
approximation. Finally we propose a generic function for the action
to cover all the mentioned scales and in addition provides a late
time acceleration for the universe. At the end we use the
observational data of the Pioneer anomalous in solar system,
rotation curve of the galaxies and Supernova Type Ia in the
cosmological scales to put constrain the parameters of the model.

 The organization of the paper is as follows: In Sec.
\ref{metric} we introduce the modified gravity in metric formalism.
Using an ansatz for the derivative of the action, we solve the field
equation for the spherically symmetric metric and derive the
dynamics in solar system and galactic scales. We use the
observational data in the solar system as well as the flat rotation
curve of the spiral galaxies to constrain the parameters of the
model. In Sec. \ref{propose} we propose a generic action where in
the small scales it reduces to the appropriate actions in the
galactic and solar system scales. SNIa gold sample data provides a
compatible results with the other observations. The conclusion is
presented in Sec. \ref{conclusion}.

\section{Spherically symmetric space}
\label{metric} Let us take an action for the gravity depends only on
the Ricci scalar as $f(R)$ where in the simple case of $f(R) =R$, it
is so-called the Einstein-Hilbert action. For a generic $f(R)$,
there are two main approaches to extract the field equations. The
first one is so-called "metric formalism" in which the variation of
action is performed with respect to the metric. In the second
approach, "Palatini formalism", the connection and metric are
considered independent of each other and we do variation for those
two parameters independently. In this work we will follow the metric
formalism.

A generic form of the action depending on the Ricci scalar can be
written as follows:
\begin{equation}
S = \frac{1}{2\kappa}\int d^4x\sqrt{-g}f(R) +S_m.
\label{generalized}
\end{equation}
Varying the action with respect to the metric results in the field
equations as:
\begin{equation}
F(R)R_{\mu \nu}-\frac{1}{2}f(R)g_{\mu
\nu}-(\nabla_{\mu}\nabla_{\nu}-g_{\mu \nu}\Box ) F(R)=\kappa T_{\mu
\nu}, \label{field}
\end{equation}
where $F=df/dR$ and $\Box\equiv\nabla_{\alpha}\nabla^{\alpha}$. From
equation (\ref{field}), we take the trace and obtain the action in
terms of $F$ and Ricci scalar as:
\begin{equation}
f(R) = \frac{1}{2}(3\Box F+FR-\kappa T). \label{trace}
\end{equation}
By taking derivative from Eq.(\ref{trace}) with respect to $r$
(radial coordinate of the metric) we rewrite this equation in terms
of $F$ and $R$ as follows:
\begin{equation}
RF'-FR'+3(\Box F)'=\kappa T',\label{04}
\end{equation}
where $' \equiv d/dr$ and $f'(R) = F(R) R'$.

Replacing $f(R)$ in favor of $F(R)$, we obtain the field equation in
terms of $F(R)$:
\begin{eqnarray}
R_{\mu\nu}-\frac{1}{4}g_{\mu\nu}R&=&\frac{\kappa}{F}(T_{\mu\nu}-\frac14g_{\mu\nu}T)\nonumber
\\ &~&+\frac{1}{F}(\nabla_\mu\nabla_\nu F-\frac{1}{4}g_{\mu\nu}
\Box F).\label{palfe2}
\end{eqnarray}

Following the method introduced in \cite{Multa}, we solve the
time-independent spherical symmetric field equation in the vacuum.
Let us take a generic spherically symmetric metric as:
\begin{equation}
 ds^2 = -B(r)\,dt^2+A(r)\,dr^2+r^2d\theta^2 +r^2\sin^2\theta
 d\phi^2.
\label{gmetric}
\end{equation}
Since the metric depends only on $r$, one can view Eq.
(\ref{palfe2}) as a set of differential equations for $F(r)$, $B(r)$
and $A(r)$. For the spherically symmetric space both sides of Eq.
(\ref{palfe2}) are diagonal and we have two independent equation. We
rewrite Eq. (\ref{palfe2}) as:
\begin{equation}
K_{[\mu]}=\frac{FR_{\mu\mu}-\nabla_\mu\nabla_\mu F-\kappa
T_{\mu\mu}}{g_{\mu\mu}}, \label{comb}
\end{equation}
where $K_{[\mu]}$ is an index independent parameter and
$K_{[\mu]}-K_{[\nu]}=0$ for all $\mu$ and $\nu$. For the vacuum
space $T_{\mu\nu}=0$, $K_{[t]}-K_{[r]}=0$ results in:
\begin{equation}
2F\frac{X'}{X}+rF'\frac{X'}{X}-2rF''=0,\label{t-r}
\end{equation}
where $X(r)=B(r)A(r)$. For $K_{[t]}-K_{[\theta]}=0$:
\begin{equation}
B''+(\frac{F'}{F}-\frac{1}{2}\frac{X'}{X})B'
-\frac{2}{r}(\frac{F'}{F}-\frac{1}{2}\frac{X'}{X})B-\frac{2}{r^2}B+\frac{2}{r^2}X=0.\label{r-th}
\end{equation}
In the case of Einstein-Hilbert action $(F=1)$ equation (\ref{t-r})
reduces to $X=1$ and equation (\ref{r-th}) reduces to the
Schwarzschild solution. We note that for this case equation
(\ref{04}) also reduces to $0=0$ identity. In generic case having a
$F$ as a function of distance or as a function of Ricci scalar, we
can obtain the metric elements from equations (\ref{t-r}) and
(\ref{r-th}).

Here we take an ansatz of $F(r)=(1+r/d)^{-\alpha}$ for the
derivative of action as a function of distance from the center,
where $\alpha$ is a small dimensionless constant $(\alpha\ll 1)$ and
$d$ is a characteristic length scale in the order of galactic size.
Similar to the case of $F=1$ we use equations (\ref{t-r}) and
(\ref{r-th}) to derive $X$ and $A$.
We start with the Eq. (\ref{t-r}), the solution results in:
\begin{equation}
X(r)=X_{0}(1+\frac{r}{d})^{-2(1+\alpha)}
(1+\frac{2-\alpha}{2}\frac{r}{d})^{\frac{4(1+\alpha)}{2-\alpha}},\label{Xgen}
\end{equation}
where $X_{0}$ is constant of integration and for $\alpha=0$ we
recover Schwarzschild metric, implies  $X_{0}=1$

In what follows we obtain metric element $B(r)$ by solving the
differential equation of (\ref{r-th}) for the solar system scales
$(r\ll d)$ and galactic scales $(r>d)$ . Once we derive the metric,
the Ricci scalar and the corresponding action can be obtained.
Finally we will use equation (\ref{04}) to check the consistency of
the solution.

\subsection{Solar system scale $(r \ll d)$}
In 1998 Anderson et al. \cite{And1} reported an unmodeled constant
acceleration towards the Sun of about $a_P=8.5\times10^{-10} m/s^2$
for the spacecrafts Pioneer 10 (launched 2 March 1972), Pioneer 11
(launched 4 December 1973), Galileo (launched 18 October 1989) and
Ulysses (launched 6 October 1990). In a subsequent report
\cite{And2} they discussed in detail many suggested explanations for
the effect and gave the value $a_P =(8.74\pm1.33)\times10^{-10}
m/s^2$ directed towards the Sun. The data covered many years staring
in 1980 when due to the large distance (20 AU) of Pioneer 10 from
the Sun the solar radiation pressure became sufficiently small. The
data was collected up to 1990 for Pioneer 11 (30 AU) and up to 1998
(70 AU) for Pioneer 10. In this section our aim is to explain this
extra acceleration by the modified gravity model.

We assume the range of $r\ll d$ in our concern and neglect all the
higher terms of $r/d$ and $\alpha r/d$. $F(r)$ in this regime
reduces to:
\begin{equation}
F(r) =  1-\frac{\alpha}{d}r.\label{F1}
\end{equation}
This expression is similar to adding the first order of the
perturbation of the action around the Einstein-Hilbert action. From
equation (\ref{Xgen}), expanding $X$ up to the first order results
in $X=X_{0}=1$.

Using (\ref{F1}) in the differential equation (\ref{r-th}) we obtain
$B(r)$ as follows:
\begin{eqnarray}
B(r)&=&\big[1+\frac{\alpha}{d}r+(\frac{3}{2}+\ln\big|\frac{\alpha}{d}-\frac{1}{r}\big|)
\frac{\alpha^2}{d^2}r^2\big]+c_1r^2\nonumber\\&~&+c_2\big[\frac{1}{3r}+\frac{\alpha}{2d}
+\frac{\alpha^2}{d^2}r+\frac{\alpha^3}{d^3}r^2\ln\big|\frac{\alpha}{d}-\frac{1}{r}\big|
 \big],\label{bb}
\end{eqnarray}
where $c_1$ and $c_2$ are the constants of the integration. For the
case of $ \alpha = 0$ (Einstein-Hilbert action) we use the
Schwarzschild metric as the zero order which implies $c_1 = 0$ and
$c_2 = -6m$. Using equation (\ref{bb}) we can obtain the Ricci
scalar of this metric. We keep up to the first order of perturbation
in metric as:
\begin{equation}
B(r)=1-\frac{2m}{r}+\frac{\alpha}{d}r,
\end{equation}
The Ricci scalar in the spherically symmetric space for a generic
case of $X$ is:
\begin{equation}
R=-\frac{1}{X}\left[B''+\frac{4}{r}B'+\frac{2}{r^2}B
-\frac{X'}{X}(\frac{1}{2}B'+\frac{2}{r}B)\right]+\frac{2}{r^2},\label{06}
\end{equation}
where substituting the metric elements, the corresponding Ricci
scalar up to the first order obtain as:
\begin{equation}
R(r)=-\frac{6\alpha}{rd},\label{Rmetric}
\end{equation}
Now to check the consistency condition, we apply the metric element
as well as the Ricci scalar in equation (\ref{04}). Here $\Box F$ up
to the first order reduce to:
\begin{eqnarray}
\Box F&=&\frac{B}{X}(F''+\frac{2}{r}F'-\frac{1}{2}\frac{X'}{X}F'
+\frac{B'}{B}F').\label{BoxF} \\
&=&-\frac{2\alpha}{rd}.\label{ssBoxF}
\end{eqnarray}
Doing Simple algebra shows the consistency of the equation for the
trace equation up to the first order of perturbation.

Now we replace $r$ in favor of $R(r)$ in equation (\ref{F1}) and
obtain $F(R)$ in terms of Ricci scalar as:
\begin{equation}
F(R)=1+\frac{6\alpha^2}{Rd^2}.\label{FR11}
\end{equation}
Finally integrating (\ref{FR11}) yields action as follows:
\begin{equation}
f(R)=R+R_0\ln\frac{R}{R_c},\label{FR12}
\end{equation}
where $R_0=6\alpha^2/d^2$ and $R_c$ is the constant of integration.

The equation of motion for a test particle from the metric can be
obtained. Using the week field regime, we define an effective
potential as:
\begin{equation}
\phi_N = -\frac{m}{r} + \frac{\alpha }{2d}r,
\end{equation}
where the acceleration of the particles from this potential is
\begin{equation}
a =-\frac{m}{r^2} -\frac{\alpha }{2d}. \label{accp}
\end{equation}
The first term at the right hand side of this equation is the
standard Newtonian gravity, however the second term is a constant
acceleration, independent of the mass. We may correspond this extra
term to the Pioneer anomalous and constrain it with the observed
value of $a_P =(8.74\pm1.33)\times10^{-10} m/s^2$ which results in
$\alpha/d\simeq 10^{-26} m^{-1}$.

\subsection{Galactic scale $(r > d)$}
Recently some of the authors have been tried to explain the dynamics
of galaxies by the modified gravity instead of assuming a dark
matter halo for the galaxy \cite{capo07}. We follow the same method
to extract the action with the ansatz of $F(r)=(1+r/d)^{-\alpha}$
proposed in this work. We assume $r$ to be larger than the
characteristic length scale of the model $d$ and write
$F(r)=(1+r/d)^{-\alpha}$ for $\alpha\ll1$ as follows:
\begin{equation}
F(r)\simeq (r/d)^{-\alpha}\simeq 1-\alpha\ln(r/d).\label{F2}
\end{equation}
This action can be considered as perturbation around the
Einstein-Hilbert action. From Eq.(\ref{Xgen}),
\begin{eqnarray}
X(r)=(\frac{r}{d})^\alpha.\label{Xup}
\end{eqnarray}
We follow the same procedure as we did in the case of solar system
to extract the metric. Using equations (\ref{F2}) and (\ref{Xup}) in
(\ref{r-th}) we obtain $B(r)$ as:
\begin{eqnarray}
B(r)&=&(\frac{r}{d})^{\alpha}\left[\frac{1}{1-\alpha}+e_1r^{-(1-\alpha/2)}
+e_2r^{2(1-\alpha/2)}\right],\nonumber\\
&=&\frac{1}{1-\alpha}(\frac{r}{d})^{\alpha}\left[1+e'_1r^{-(1-\alpha/2)}
+e'_2r^{2(1-\alpha/2)}\right],\label{BAup}
\end{eqnarray}
where $e_{i}$'s are the constants of integration and
$e'_{i}=e_{i}(1-\alpha)$ for $i=1,2$. For $ \alpha = 0$ equations
(\ref{F2}) and (\ref{Xup}) reduce to $F=1$ and $X=1$ and we expect
to recover Schwarzschild-de Sitter metric which yields $e'_1 =-2m$
and $e'_2=\frac{1}{12}\Lambda$, where $\Lambda$ is the cosmological
constant. For generic case when $\alpha \neq 0$, from the
dimensional analysis, the constants of the integration obtain as
$e'_1 =-(2m)^{1-\alpha/2}$ and $e'_2 = (\Lambda/12)^{1-\alpha/2}$.
We rewrite the metric elements after fixing $e'_i$'s:
\begin{eqnarray}
B(r)&=&\frac{1}{1-\alpha}\left[1-(\frac{2m}{r})^{1-\alpha/2}
+(\frac{\Lambda r^2}{12})^{(1-\alpha/2)} \right](\frac{r}{d})^{\alpha},\nonumber\\
A(r)&=&(1-\alpha)\left[1-(\frac{2m}{r})^{1-\alpha/2} +(\frac{\Lambda
r^2}{12})^{(1-\alpha/2)} \right]^{-1}. \label{BAup2}
\end{eqnarray}
From the metric elements we get the following Ricci scalar
\begin{eqnarray}
R(r)&=&-\frac{1}{(1-\alpha)r^2}[3\alpha+(12-3\alpha)(\frac{\Lambda
r^2}{12})^{1-\alpha/2}\nonumber \\
&~&~~~~~~~~~~~~~~~~~~~+\frac{\alpha^2}{2}(1-
3(\frac{2m}{r})^{1-\alpha/2})].\label{R044}
\end{eqnarray}
We keep Ricci scalar up to the first order term in $\alpha$ and
$\Lambda$, then (\ref{R044}) reduces to
\begin{equation}
R(r)=-\frac{3\alpha}{r^2}-\Lambda.\label{R045}
\end{equation}
Again to check the consistency of the solution in this space, we
substitute (\ref{F2}) and (\ref{R045}) in equation (\ref{04}). The
solution of this space satisfies the trace equation up to the first
order of perturbation in terms of $\alpha$. We note that in general,
for non-perturbed case the solutions might be inconsistent. Here the
solution is valid only up to the first order of perturbation.

Eliminating $r$ in favor of $R$ from equation (\ref{R045}) and using
(\ref{F2}), the derivative of action obtain as:
\begin{equation}
F(R)=(\frac{d^2}{3\alpha}|R+\Lambda|)^{\alpha/2},\label{FR21}
\end{equation}
then
\begin{equation}
f(R)=\frac{1}{1+\alpha/2}(\frac{d^2}{3\alpha})^{\alpha/2}
|R+\Lambda|^{1+\alpha/2}.\label{FR22}
\end{equation}
For simplicity let us write action as:
\begin{equation}
f(R)=f_0|R+\Lambda|^{1+\alpha/2}.\label{f22}
\end{equation}

\begin{figure}
\epsfxsize=8.truecm\epsfbox{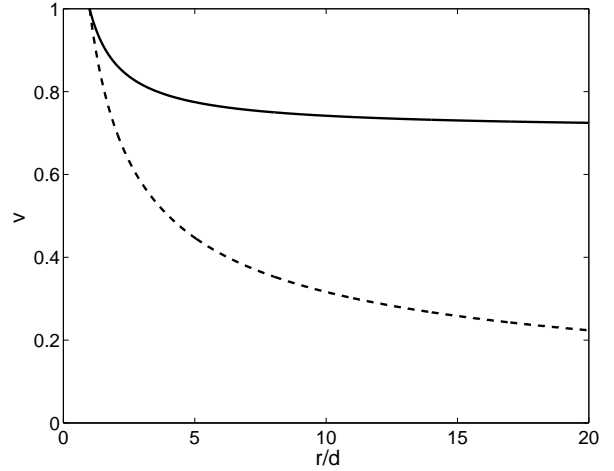} \narrowtext
\caption{Comparing the rotation curve of galaxy in Newtonian
gravity, $v\propto 1/r$ (dashed-line) with the rotation curve from
the modified gravity in the weak filed regime (solid-line), see
equation (\ref{vv}). The parameters of galaxy is taken as $M =
10^{11} M_\odot$ (mass of galaxy), $\alpha \simeq 10^{-6}$ and
$d\simeq 10$ kpc.} \label{f1}
\end{figure}

The dynamics of a test particle around this metric follows the
geodesic equation in weak field regime,
\begin{equation}
\ddot{r} + \Gamma^{r}{}_{tt} = 0,
\end{equation}
where substituting the corresponding metric elements we get the
following velocity for a particle rotating around the center of
galaxy:
\begin{equation}
v =
\frac{c}{\sqrt{2}}(\frac{r}{d})^{\alpha/2}\left[(\frac{2m}{r})^{1-\alpha/2}
+ \alpha\right]^{1/2}, \label{vv}
\end{equation}
where we ignored $\Lambda r^2$ term as it is five order of magnitude
smaller than $2m/r$. For $\alpha=0$ we recover the standard
Newtonian law for the rotation velocity of a test particle, in which
$m = GM/c^2$ and $M$ is the mass of galaxy. The extra term in Eq.
(\ref{vv}) may provide contribution to the flat rotation curve.
Figure (\ref{f1}) compares the rotation curve of a test particle
around the center of galaxy with an arbitrary units in the modified
and standard Newtonian gravity. Here we model the mass of galaxy
spherically distributed up to $3.3$ kpc and obtain the rotation
curve up to $66$ kpc. For a typical spiral galaxy with the mass of
$M = 10^{11} M_\odot$ and at the large distances (e.g. $r> d$) from
the center, $v \sim 200~km~s^{-1}$ which roughly constrain $\alpha
\simeq 10^{-6}$. In the previous section we had an estimation for
$\alpha/d \simeq 10^{-26}~m^{-1}$ which provides the characteristic
length scale of the model, $d \simeq 10~kpc$. We note that while
equation (\ref{vv}) provides a flat rotation curve for the galaxy
but it doesn't support the Tully-Fisher relation.

\section{proposing a generic action}
\label{propose}
 In the previous section we obtained the asymptotic behavior of
an action in the galactic and solar system scales. Those two actions
could describe the observations in the corresponding length scales
without need to a dark matter. The action for the small scale vary
with a logarithmic function and for the galactic scales with a
power-law function. This asymptotic behavior of the actions guides
us to guess a generic action which can cover also those two scales.
Here we propose the action of
\begin{equation}
f(R) = R+\Lambda + \frac{R+\Lambda}{R/R_0
+2/\alpha}\ln{\frac{R+\Lambda}{R_c}}.\label{hybrid}
\end{equation}
For the range of $R\gg \Lambda$ and $R/R_0\gg 2/\alpha$, action
reduces to
\begin{equation}
f(R) = R + R_0\ln({\frac{R}{R_c}}).
\end{equation}
Comparing with (\ref{FR12}) provides $R_0 = 6\alpha^2/d^2$. Using
$|R(r)| = {6\alpha}/{rd}$ and $R/R_0\gg 1/\alpha$, satisfies the
solar system range of $r\ll d$.

On the other hand for $\alpha\ll 1$ and $R\simeq R_0\simeq \Lambda$
action (\ref{hybrid}) can be written as:
\begin{equation}
f(R)=(R+\Lambda)\left[1 +
\frac{\alpha}{2}\ln(\frac{R+\Lambda}{R_c})\right],
 \label{cosmfr}
\end{equation}
where for small $\alpha$, we write action as:
\begin{equation}
f(R) = \frac{(R+\Lambda)^{1+\alpha/2}}{{R_c}^{\alpha/2}}.
\label{cosf}
\end{equation}
For $\alpha\ll 1$, the action reduces to $f(R) = R + \Lambda$. We
expect the best parameters of the model from SNIa and CMB
experiments should be around the $\Lambda$CDM model. To see the
consistency of the this action with the matter dominant epoch, we
let $R\gg \Lambda$ and $R\gg R_0$. In this case the action reduces
to $f(R)\rightarrow R$ (i.e. Einstein Hilbert action) and the scale
factor changes as $a\propto t^{2/3}$ with time.

In what follows we put constrain on the parameters of the model in
(\ref{cosmfr}). The generic FRW equation in modified gravity is
\begin{equation}
3H\dot F+ 3H^{2}F-\frac{1}{2}(f-RF)=\kappa\rho_{m}. \label{mfrw}
\end{equation}
We use SNIa gold sample with the prior of flat universe to constrain
$\Omega_m$, $\Omega_\Lambda$ and $\alpha$. Using action of
(\ref{cosmfr}), equation (\ref{mfrw}) is written as follows:
\begin{eqnarray}
&&H^2-\frac{\Lambda}{6}+\frac{\alpha}{2}\{H^2[\frac{R}{R+\Lambda}
+\ln\frac{R+\Lambda}{R_c}]+\frac{R^2}{R+\Lambda}\nonumber\\
&&~~~~~~~~~~~~~~~~~~~~~~~~~~+\frac{R+2\Lambda}{(R+\Lambda)^2}H\dot{R}\}
=H_0^2\Omega_m a^{-3}, \label{diffeq}
\end{eqnarray}
where $3H_0^2\Omega_m=\kappa\rho_m$. For $\alpha = 0$ we recover
standard FRW equation.
\begin{equation}
H
=H_0(\Omega_ma^{-3}+\Omega_{\Lambda})^{1/2}, \label{lcdm}
\end{equation}
 On the other hand variation
of action we respect to the metric guarantee the conservation of
energy momentum and for the matter we can write $\rho = \rho_0
a^{-3}$.

From the constrain of rotation curve of the spiral galaxies in the
previous section, $\alpha \simeq 10^{-6}$, we assume $\alpha\ll 1$
and this term is considered as a perturbation parameter in equation
(\ref{diffeq}). We solve equation (\ref{diffeq}) by perturbing the
Hubble parameter around $\Lambda$CDM solution, $H=H^{(0)}+\alpha
H^{(1)}$, in which $H^{(0)}$ obtain from equation (\ref{lcdm}) and
$H^{(1)}$ is calculated from
\begin{eqnarray}
&&H^{(1)}=-\frac{1}{4}\frac{H^{(0)}}{R^{(0)}+\Lambda}\{H^{(0)}R^{(0)}
+\frac{R^{(0)}+2\Lambda}{R^{(0)}+\Lambda} \dot{R}^{(0)}
+\frac{{R^{(0)}}^2}{6H^{(0)}}\nonumber\\
&&~~~~~~~~~~~~~~~~~~~~~~~+H^{(0)}(R^{(0)}+\Lambda)^3\ln\frac{R^{(0)}+\Lambda}{R_c}
\},\label{H1}
\end{eqnarray}
where $R^{(0)}$ and $\dot{R}^{(0)}$ are the zero order terms obtain
from $H^{(0)}$ and $\dot{H}^{(0)}$. The relevant parameter for
comparing the theoretical model with SNIa data is the luminosity
distance $D_L=D_L(z;\Omega_m,\alpha,R_c,\Lambda,h)$ and is related
to the distance modulus of the supernovas as follows:
\begin{figure}
\epsfxsize=9.7truecm\epsfbox{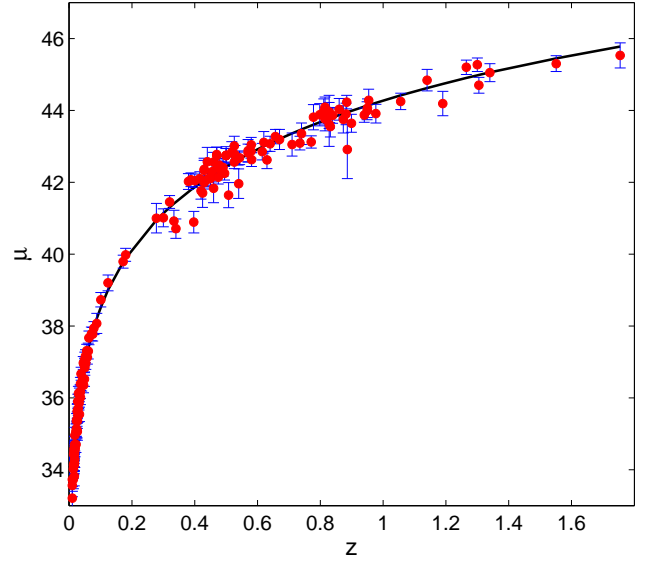} \narrowtext \caption{ Distance
modulus of the SNIa new Gold sample in terms of redshift. The solid
line shows the best fit values with the corresponding parameters of
$h=0.64$, $\Omega_m=0.31$, $\Omega_{\Lambda}=0.69$ and $\alpha\ll
10^{-3}$ with the corresponding $\chi^2_{min}/N_{d.o.f} =1.14$. }
\label{f2}
\end{figure}
\begin{eqnarray} \label{e8}
&&\mu= m-M= 5\log_{10}\left[ \frac{D_L}
{10~pc}\right], \nonumber \\
&&D_L = c(1+z)\int_0^z
\frac{dz}{H(z;\Omega_m,\alpha,R_c,\Lambda,h)},
\end{eqnarray}
where the K-correction is included in the distance modulus of the
supernovas.

We do likelihood analysis letting the Hubble parameter $h$, the
cosmological parameters $\Omega_m$ and $\Omega_\Lambda$, and
$\alpha$ as the free parameters to find the best values. The
comparison between the observed and theoretical distance modulus is
done by $\chi^2$ fitting as follows:
\begin{eqnarray}\label{chi_sn}
\chi^2=\sum_{i}\frac{[\mu_{obs}(z_i)-\mu_{th}(z_i;\Omega_m,\alpha,R_c,\Lambda,h)]^2}{\sigma_i^2},
\end{eqnarray}
The best value for $\chi^2$, normalized to the number of degree of
freedom is $\chi^2/N_{d.o.f} = 1.14$ . The corresponding best values
for the parameters of the model is: $\Omega_m=0.31$,
$\Omega_{\Lambda}=0.69$, $h=0.64$ and  $ \alpha \ll  10^{-3}$. The
constrain on $\alpha$ is consistent with the results from the
rotational velocity of spiral galaxies, $\alpha\simeq 10^{-6}$.
Finally we should point out that $R_c$ is not sensitive to the SNIa
data.


\section{Summary and Discussion}
\label{conclusion} In this work we tried to explain the anomalous in
the acceleration of the Pioneer spacecraft and flat rotation curve
of spiral galaxies in the framework of the modification of the
gravity. We started by assuming an ansatz for the derivative of
action in terms of distance from the center and did the inverse
procedure to derive the metric and action of the space. In the solar
system scale we extract a logarithmic extra term to the
Einstein-Hilbert action and in the galactic scale we follow the same
procedure and found a power law action. The solution in both two
regimes obtained as perturbation around the Einstein-Hilbert action
and we showed that within this approximation the solutions are
consistent with the modified gravity equations. We note that in
generic case we may not find a consistent solution in the
spherically space \cite{saf07}. Finally we proposed a generic action
where in the asymptotic regimes reduce to our desired metric and
actions in the solar system and galactic scales. For the
cosmological scales this action provides a late time acceleration
for the universe. Finally we used the pioneer data, flat rotation
curve of galaxies and the CMB and Supernova Type Ia gold sample to
put constrain on the parameters of the model.

{\bf Acknowledgements} We would like to thank anonymous referee for
useful comments.

\end{document}